\documentclass[fleqn,usenatbib,useAMS,referee]{mnras}

\usepackage{graphicx}   
\usepackage{amsmath}    
\usepackage{amssymb}    
\usepackage{multicol}        
\usepackage{bm}     
\usepackage{pdflscape}  

\title[SOC in type I X-ray bursts]{Self-organized criticality in type I X-ray bursts}

\author[Wang, Wang \&Dai]{J. S. Wang$^{1,2}$, F. Y. Wang$^{1,3}$\thanks{Contact e-mail: FYW(fayinwang@nju.edu.cn) or JSW(jieshuang.wang@mpi-hd.mpg.de)} and Z. G. Dai$^{1,3}$\\$^1$School of Astronomy and Space Science, Nanjing University, Nanjing 210093, China; \\
$^2$Max-Planck-Institut f\"ur Kernphysik, Saupfercheckweg 1, D-69117 Heidelberg, Germany; \\
$^{3}$Key Laboratory of Modern Astronomy and Astrophysics (Nanjing University), Ministry of Education, China}

\begin{document}

\label{firstpage}
\pagerange{\pageref{firstpage}--\pageref{lastpage}}
\maketitle

\begin{abstract}
{Type I X-ray bursts in a low-mass X-ray binary (LMXB) are caused by
unstable nuclear burning of accreted materials. Semi-analytical and
numerical studies of unstable nuclear burning have successfully
reproduced partial properties of this kind of burst. However, some
other properties (e.g. the waiting time
) are not well explained.
In this paper, we find that the probability distributions of
fluence, peak count, rise time, duration and waiting time can be
described as power-law-like distributions. This indicates that type
I X-ray bursts may be governed by a self-organized criticality (SOC)
process. The power-law index of waiting time distribution (WTD) is
around $-1$, which is not predicted by any current waiting time
model. We propose a physical burst rate model, in which the mean
occurrence rate is inversely proportional to time $\lambda\propto
t^{-1}$. In this case, the WTD is well explained by a
non-stationary Poisson process within the SOC theory. In this
theory, the burst size is also predicted to follow a power-law
distribution, which requires that the emission area
possesses only part of the neutron star surface.
Furthermore, we find that the WTDs of some astrophysical phenomena
can also be described by similar occurrence rate models.}
\end{abstract}


\maketitle

\section{Introduction}
Type I X-ray bursts or thermonuclear bursts are nuclear shell
flashes in low-mass X-ray binary (LMXB) systems. The accreted
hydrogen/helium matter accumulates into a thin shell upon the
neutron star surface and releases nuclear energy unstably
\citep{Woosley1976,Maraschi1977,Joss1977}. Recently, a class of
`superburst' was also found and though to be caused by unstable
burning carbon \citep{Strohmayer2002}.  Thermonuclear bursts were
first observed in 1970's \citep{Grindlay1976,Belian1976}.
Since then, type I X-ray bursts have been widely studied
observationally and theoretically.

This type of bursts offer a chance to probe the properties of
neutron stars, such as the mass-radius relation, magnetic field, and
spin \citep{Lewin1993,Watts2012}. Theoretical models have been
proposed and successfully explain most of burst properties
\citep{Joss1978,Fujimoto1981,Ayasli1982,
Fujimoto1987,Fushiki1987,Lewin1993,Narayan2003,Woosley2004,Fisker2008,Jose2010},
and several ignition regimes have been identified as resulting from
different burst fuel compositions and accretion rates \citep[e.g.
see][]{Fujimoto1981,Narayan2003}. However, some properties, such as
waiting time or recurrence time, have not been fully understood.

For normal bursts (H/He flashes), the theoretical waiting time is
$\Delta t=\Delta M_{\rm{crit}}/ \dot{m}$, roughly a few hours in
LMXBs, where $\Delta M_{\rm{crit}}$ is the critical shell mass when
burst can be ignited and $\dot{m}$ is the accretion rate \citep[e.g.
see][]{Narayan2003}. \cite{Galloway2004} found that the waiting time
for GS 1826-24 is proportional to $\dot{m}^{-1.05\pm0.02}$, assuming
that the accretion rate $\dot{m}$ is linearly proportional to the
observed X-ray flux. However, this should be not important. Because,
for most sources the accretion rate changes only in a small range
\citep[e.g. see Table\,10 of][]{Galloway2008}. \cite{Keek2010}
studied the effect of the data gaps using Monte Carlo simulations.
They assumed that the ``intrinsic'' waiting time distribution of EXO
0748-676 follows a bimodal distribution of mean values at 12.7
minutes and 3.0 hours with 16\% widths. Then due to the effect of
data gaps, the resulting waiting time will span 2-3 orders of
magnitude. But it should be noted that further interpretation is
needed that why the ``intrinsic'' waiting time follows a bimodal
distribution. Even though, this still does not cover the observed
waiting times well. A detailed study of normal bursts from
\emph{Rossi~X-ray~Timing~Explorer} (RXTE) showed that the waiting
times from single source can span 4-5 orders of magnitude, from a
few minutes to years \citep{Galloway2008}. In their paper,  the
`superburst' (carbon flashes) is not taken into consideration. In
addition, several challenges to the theoretical models of the
waiting time were also noted \citep{Galloway2008,Keek2010},
including very short waiting-time bursts. Meanwhile, it is not fully
understood what causes the asymmetric brightness patches, also known
as burst oscillations, which might be related to the emission area
of the bursts \citep{Strohmayer1996,Watts2012}. Thus further study
is needed.

In this paper, we analyze the type I X-ray burst catalog of
\cite{Galloway2008} using a statistical method. We focus on the
probability distributions of waiting time, burst fluence, peak
counts, rise time and duration time. Below, we will explain waiting
time distribution (WTD) by a non-stationary Poisson process, where
the mean burst rate $\lambda\propto \dot{m}/\Delta M$ when $\Delta
M>\Delta M_{\rm{crit}}$. This is quite a bit different with previous
understanding of waiting time $\Delta t=\Delta M_{\rm{crit}}/
\dot{m}$. Data analysis and results are presented in the next
section. In section 3, we extend our model of waiting time
distribution. Summary and discussions are presented in section 4.

\section {Data analysis and results}
There are three burst cases based on the accretion rates and
ignition regimes. The burst property is quite different for them. We
give a brief summary of these cases in Table 1. More detailed
summary can be seen in Table\,1 of \cite{Galloway2008} and
references therein. In this paper, we study the catalogue
compiled by \cite{Galloway2008}. This catalogue contains 1187 burst
from 48 accreting neutron stars, while only four of them have around
100 bursts \citep[see Table\,4 of][]{Galloway2008}. Therefore, we
focus on these four sources and list some properties of them in
Table 1. The largest sample, 4U 1636-536, contains 171 bursts.
However, this source has both pure He bursts and mixed H and He
bursts. Since the burst properties (e.g. peak counts, duration, rise
time, and fluence) differ a lot in these two burst cases, we do not
choose it as an example for study. Instead, we take the second
largest source 4U 1728-34 as an example, which contains only pure He
flashes.

\begin{figure}
\centering
\includegraphics[width=0.8\textwidth]{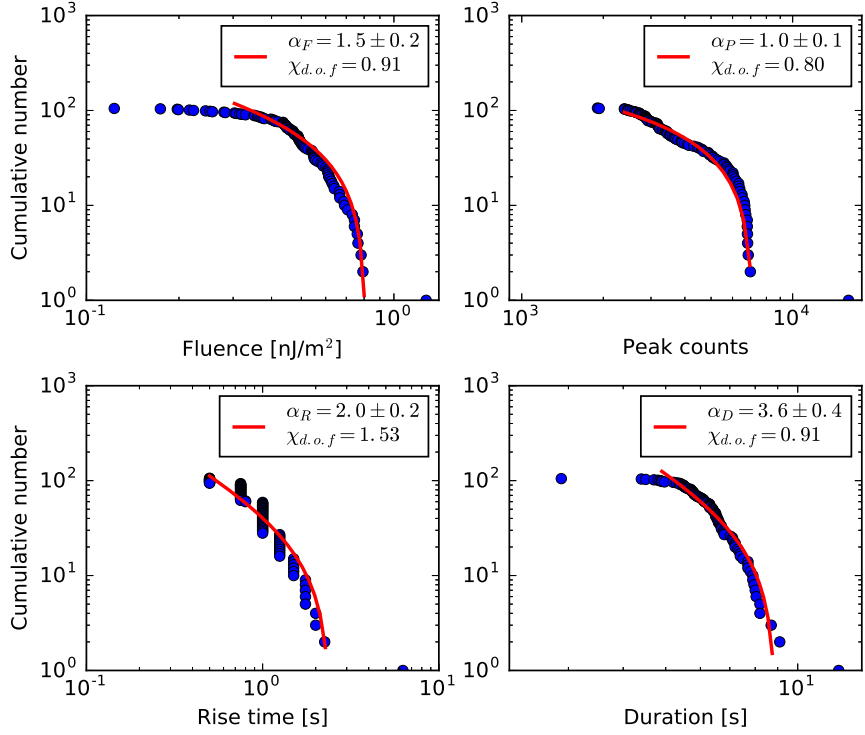}
\caption{Cumulative distributions of  fluence, peak counts, rise
time and duration for 4U 1728-34. The best-fitted results are shown
by the red lines. The best-fitted parameters are given in each
panel. \label{4U_1728_34}}
\end{figure}

Since the burst number of this source is small, we study the
cumulative number distribution of the data. If the differential
distributions of the fluence, peak counts, rise time and duration
time follow the Pareto distributions, i. e. $N(x)\propto
x^{-\alpha}$ for $x\geq x_{\rm min}$, the cumulative number would be
\begin{eqnarray}
N_{\rm cum}(>x)=(N_{\rm tot}-1){x_{\rm max}^{1-\alpha}-x^{1-\alpha} \over x_{\rm max}^{1-\alpha}-x_{\rm min}^{1-\alpha}} +1, {~\rm when~ x\geq x_{\rm min}} ,
\end{eqnarray}
where $x$ corresponds to observed parameter and its minimum and
maximum value are $x_{\rm min}$ and $x_{\rm max}$. $N_{\rm tot}$ is
the total number of the sample. The index $\alpha$ is the scaling
parameter for the distributions, namely, $\alpha_F$ for fluence,
$\alpha_P$ for peak counts, $\alpha_R$ for rise time and $\alpha_D$
for duration. It is worth noting that a physical threshold of an
instability or the incomplete sampling of small events, which is
widely found in astrophysics \citep{Aschwanden2015}, can cause
fluctuations at the left part. Thus we do not take into account this
part when fitting.
We fit the parameter $\alpha$ by minimizing the reduced
$\chi^2$, which is
\begin{equation}
    \chi_{d.o.f} = \sqrt{ {1 \over (N_{\rm tot}-n_{\rm par})}
    \sum_{i=1}^{N_{\rm tot}} { [N_{\rm cum}(x_i) - N_{\rm cum, obs}(x_i)]^2
    \over \sigma_{\rm cum,i}^2 } }\ ,
\end{equation}
where $n_{\rm par}$ is the number of parameters in the calculation,
$N_{\rm cum, obs}(x_i)$ is the observed cumulative number of events
$x>x_i$ and $\sigma_{\rm cum,i}=\sqrt{N_{\rm cum, obs}(x_i)}$. The
best-fitted values of the scaling parameters and uncertainties are
$\alpha_F=1.5\pm0.2$, $\alpha_P=1.0\pm0.1$, $\alpha_R=2.0\pm0.2$ and
$\alpha_D=3.6\pm0.4$ for 4U 1728-34. We show the best-fitted value
of the scaling parameters and the minimum $\chi_{d.o.f}$ in
Figure\,\ref{4U_1728_34}. For the pure random noise, the expected
$\chi_{d.o.f}$ is around 1. Using the same method, we find that the
other three sources in Table\,1 also have similar distributions as
4U 1728-34. We also compare our results with the predicted
values of the fractal-diffusive transport self-organized criticality
(SOC) model proposed by \cite{Aschwanden2014}, which gives
$\alpha_F=1.5$, $\alpha_P=1.67$ and $\alpha_D=2.0$. The value of
$\alpha_P$ and $\alpha_D$ are not well consistent with the model
predictions, while $\alpha_F$ shows good agreement. But in fact, the
scaling parameters in some astrophysical phenomena have significant
deviations from this model \cite[see Table 1 of][for a good
summary]{Aschwanden2014}. One possible reason is that the sample
size is small. For example, the incomplete sampling of small events
near the threshold $x_{\rm min}$ \citep{Aschwanden2015} could make a
big difference, especially when the observed maximum value $x_{\rm
max}\sim~10x_{\rm min}$ in our research.

Based on the criteria proposed by
\cite{Aschwanden2011,Aschwanden2014}, which are statistical
independence of events, non-linear growth phase, random rise times
and power-law distribution of these observed parameters, type I
X-ray burst can be explained by a SOC process. The SOC theory
predicts that subsystems self-organize owing to some driving force
to a critical state, at which a slight perturbation can cause a
chain reaction of any size within the system \citep{Bak1987}.

The SOC theory also predicts a power-law distribution of waiting
times \citep{Aschwanden2011,Markovic2014}. The WTD has been widely
discussed in astrophysics
\citep{Wheatland1998,Aschwanden2010,Wang2013,Li2014,Wang2015,Guidorzi2015,Wang2017}.
In the SOC theory, WTD is usually described as a (non-)stationary
Poisson process. The probability function is expressed by
\citep{Wheatland1998}
\begin{equation}
P(\Delta t)=
{\int^T_0 \lambda(t)^2 e^{-\lambda(t)\Delta t} dt\over\int^T_0 \lambda(t)dt}\propto (\Delta t)^{-\alpha_{\Delta t}},
\label{pdf}
\end{equation}
where $\Delta t$ is the waiting time, $T$ is the period, and
$\lambda(t)$ is the burst occurrence rate. Below, we model the
occurrence rate $\lambda$.

We propose that the burst rate can be modelled by $\lambda(t)\propto
t^{-\gamma}$. The reason is shown below. Type I X-ray bursts are
induced by a thin-shell instability. In LMXBs, the accreted H/He
fuel accumulates on the stellar surface and forms a shell. The
typical time for nuclear energy release is $\tau_N\propto\Delta
M/L_N$, where $\Delta M$ is the mass of the shell and $L_N$ is the
nuclear energy generation rate \citep{Hoshi1968,Joss1977}. In the
quiescent stage, $L_N$ remains almost unchanged. Meanwhile, the
energy will be diffused in a typical time $\tau_d\propto\Delta M^2$,
which also depends on the stellar temperature and the opacity of the
shell \citep{Hoshi1968,Ayasli1982}. The condition for triggering a
burst is $\tau_N/\tau_d=\Delta M_{\rm{crit}}/\Delta M\leq1$, where
the critical shell mass $\Delta M_{\rm{crit}}$ is obtained by
$\tau_N=\tau_d$
 \citep{Hoshi1968,Ayasli1982}. Alternative descriptions of this critical condition
 are also used \citep{Fujimoto1981,Fujimoto1987,Narayan2003}. We assume that
$\Delta M=\Delta M_{\rm{crit}}$ happens at a typical time $t_0$. When the shell
grows thick enough at a time $T$, any perturbation affects the stellar hydrostatic
structure significantly and leads to stable nuclear burning  \citep{Schwarzschild1965}.
Therefore, a burst can be triggered only in a time range $t_0\leq t<T$.
Since these time-scales are mainly determined by the shell mass, it is plausible to
assume that the occurrence rate also mainly depends on the shell mass. Because
the accreted mass is proportional to time, we can adopt the burst rate as
\begin{equation}
\lambda(t) = w t^{-\gamma}~~~~{\rm for}~t_0\leq t<T. \label{occ}
\end{equation}
The occurrence rate is taken as $\lambda=0$ when $t<t_0$. We put all
the time-dependent effects into the power-law index $\gamma$, which
contains the dependence on shell mass $\Delta M=\dot{m} t$ and the
possible time evolution of accretion rate $\dot m(t)$. The
coefficient $w$ is a normalization constant, i.e.,
\begin{equation}
\int^T_0 \lambda(t)dt=w\int^T_{t_0} t^{-\gamma}dt=1. \label{nor}
\end{equation}
Accordingly,  there are three free parameters ($\gamma$, $t_0$ and $T$) in
this model. Substituting the above equation into equation
(\ref{pdf}), we can obtain the differential probability distribution
\begin{eqnarray}
P(\Delta t)= \int^T_{t_0} \lambda(t)^2 e^{-\lambda(t)\Delta t} dt
= w^2\int^T_{t_0} t^{-2\gamma} e^{(-wt^{-\gamma}\Delta t)}dt\;\nonumber\\
={w\over\gamma\Delta t}\int^T_{t_0}t^{-\gamma+1} e^{(-wt^{-\gamma}\Delta t)}d(-wt^{-\gamma}\Delta t).\label{pdfmod}
\end{eqnarray}

\begin{figure}
\centering
\includegraphics[width=0.8\textwidth]{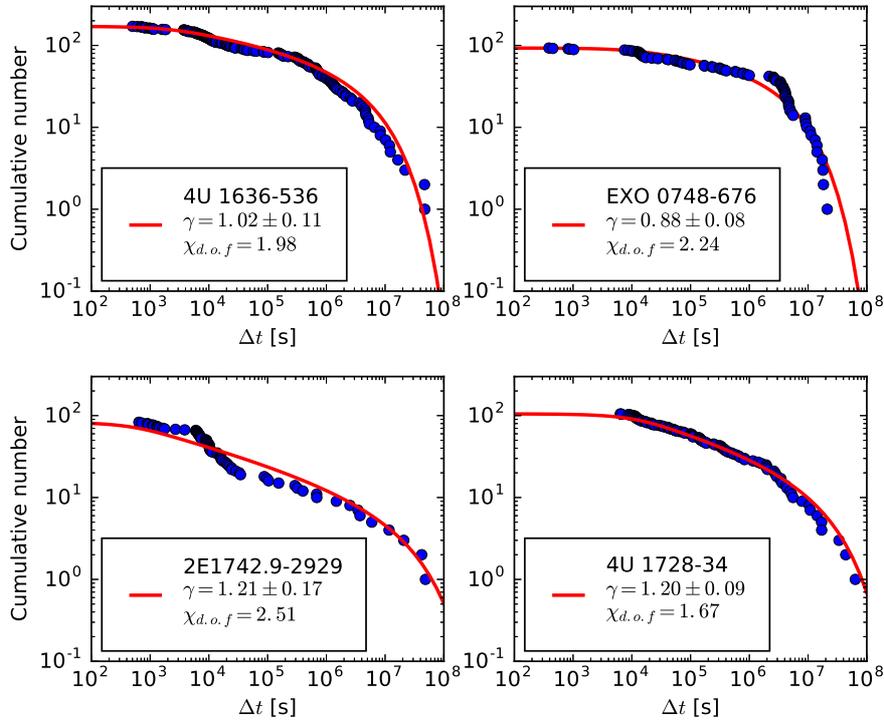}
\caption{Cumulative number distributions and fitting results
of four representative sources. The best-fitteing results are shown
as red lines. \label{wtd}}
\end{figure}

\begin{figure}
\centering
\includegraphics[width=0.8\textwidth]{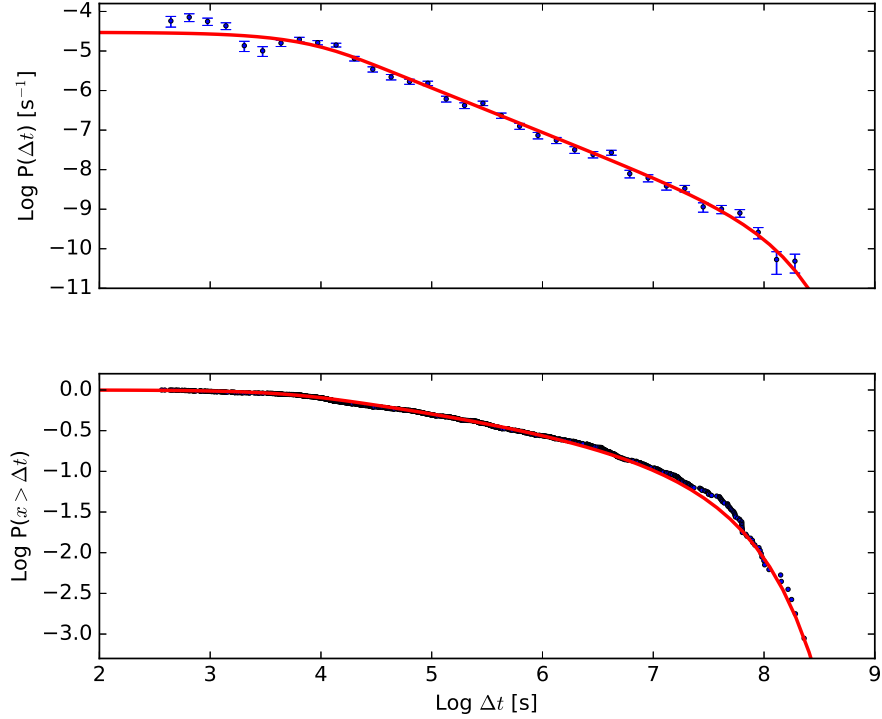}
\caption{The differential probability distribution (up
panel) and the cumulative distribution (bottom panel) of the full
sample. The best-fitted values are $\gamma=1.14\pm0.04$,
$t_0=1.15\pm0.33\times10^3$~s and $T=6.0\pm1.5\times10^6$~s.
\label{wtdfull}}
\end{figure}

The full catalogue contains 1128 waiting times, calculated in the
observer's frame \citep{Galloway2008}. In this case, the
gravitational redshift is important, which is hard to determine for
each source. But it seems that for most equations of state in
neutron stars, the gravitational redshifts span a relative narrow
range \citep{Lattimer2001}. Thus it is plausible to ignore the
gravitational redshift effect, especially when we study the WTDs in
a single source. Here we study the selected four representative
sources as shown in Table\,1 and the full sample (1128 waiting
times). The cumulative number distributions of four sources
are shown in Figure \ref{wtd}. The power-law indices $\gamma$ of
these four samples are well consistent with each other and the
typical value is $\gamma\sim-1$. We use the above waiting time model
to fit the data by the transformation of $N_{\rm
cum}=N_{tot} P(x>\Delta t)$, where $P(x>\Delta t)=\int^T_{\Delta t}
P(x) dx$. The best-fitting curves are shown in Figure \ref{wtd} and
the best-fitting parameters are given in Table 1. Then we study the
full sample. Since the number of waiting time is large enough, we
study both the differential probability distribution and cumulative
probability distribution, which are shown in Figure \ref{wtdfull}.
The best-fitting values are $\gamma=1.14\pm0.04$,
$t_0=1.15\pm0.33\times10^3$~s and $T=6.0\pm1.5\times10^6$~s. The
value of $\gamma$ is well consistent with those of four
representative samples. The typical value $\gamma\sim 1$ implies
$\lambda\propto \Delta M^{-1}\propto t^{-1}$. Analytically, we
substitute $\gamma=1$ into equation (\ref{nor}), which gives
$w=1/\ln (T/t_0)$. After substituting it into equation
(\ref{pdfmod}), the probability distribution of occurrence rate is
\begin{eqnarray}
P(\Delta t)= {1\over\ln (T/t_0)\Delta t} (e^{-\Delta t\over
T\ln(T/t_0)}- e^{-\Delta t\over t_0\ln(T/t_0)}).\label{eq:p-1}
\end{eqnarray}
In general case, the value of $T$ is much larger than $t_0$, i.e.,
$T\gg t_0$. So the term $e^{-\Delta t/T\ln (T/t_0)}- e^{-\Delta t/
t_0\ln(T/t_0)}$ has a maximum value of $e^{-t_0\ln(T/t_0)/T}-
e^{-\ln(T/t_0)}\approx e^{-t_0\ln(T/t_0)/T}$ at $t={Tt_0\ln^2(T/t_0)
\over(T-t_0)}\approx t_0\ln^2(T/t_0)$, and varies in the range
$[e^{- t_0/T\ln (T/t_0)}- e^{- 1/\ln(T/t_0)},e^{-t_0\ln(T/t_0)/T}]$
For typical values of $T=10^6$\,s and $t_0=100$\,s, this range
becomes $[0.1,1]$. Thus this part is almost changeless. Then the
probability distribution of occurrence rate is approximately
$P(\Delta t)\propto \Delta t^{-1}$, which is consistent with the
data. However, the fractal-diffusive transport SOC model
proposed by \cite{Aschwanden2014} predicts that $\alpha_D =
\alpha_{\Delta t}=2$. There is no model that predicts a
$\alpha_{\Delta t}=1$ power-law index of WTD.

As shown in Table\,1, the time scales $t_0$ and $T$ differs a little
in four sources, because these time scales depend on the accretion
rate and fuel compositions. Theoretically, if the accretion rate and
the composition of accreted material are changeless, the time scales
$t_0$ and $T$ for single sources would be almost constant. Thus, we
have almost pure power-law distribution in 4U 1728-34, as shown in
Figure \ref{wtd}. If the accretion rate and composition change,
these time scales would evolve. So only their mean value can be
obtained statistically.
Meanwhile, we should keep in mind that the data gaps of RXTE
observations could introduce an uncertainty on $t_0$. The main data
gaps originate from the South Atlantic Anomaly and the Earth
occultation, which pollute the data with waiting times around
$0.5-2$ satellite period, about $0.28-1.1\times10^4$\,s
\citep{Keek2010,Aschwanden2010}, which is close to $t_0$.

Since SOC theory is well consistent with the statistical properties
of type I X-ray burst, it is plausible to infer that the other `SOC
parameter' such as the burst size obeys a power-law-like
distribution based on the macroscopic description of SOC systems by
\cite{Aschwanden2014}. In their model, the power-law-like
distributions of peak flux, energy and duration are derived based on
the condition that the size of the event follows a power-law-like
distribution. Actually, the power-law-like distribution of size have
been observed in many astrophysical phenomena, such as lunar
craters, asteroid belt, and various solar phenomena \citep[see][and
references therein]{Aschwanden2011,Aschwanden2014,Aschwanden2016}.
In the same case, we would also expect that the size (e.g. depth,
area, and volume) of type I X-ray bursts obeys a power-law-like
distribution. In this case, it is very possible that not all the
accreted materiel burns during a burst. Some materiel will remain,
and thus it take less time to trigger the next burst.
\cite{Gottwald1987} suggested that 10-15 percent of the fuel could
remain to the subsequent burst by studying the bursts of EXO
0748-676. But more studied are needed to obtain the amount of
unburned material in a burst. Recently, \cite{Keek2017} found that
the short waiting time burst can be explained if some fuel is left
unburned during a burst at a shallow depth by using one-dimensional
simulations. It is also possible that the emission area is just a
part of the surface,  such as a hot spot.  And such hot spot model
is used to explain the burst oscillations
\citep{Strohmayer1996,Watts2012}. Thus more detailed research for
the amount of burst fuel is needed, which is beyond the scope of
this paper.

\section{Extension of the WTD model}

Since we have successfully used this WTD model to explain type I
X-ray burst, we further investigate the applicability of this model.
Generally, a function $f(\lambda)d\lambda=dt/T$ is adopted to fit
waiting time distribution, where $f(\lambda)\propto\lambda^{-p}
e^{-\beta \lambda}$ with two parameters of $p$ and $\beta$
\citep{Aschwanden2011,Li2014,Guidorzi2015}. The resulting
probability function can be expressed as $P(\Delta
t)=(2-p)\beta^{2-p}(\beta+\Delta t)^{p-3}$ with $0\leq p<2$
\citep[equation 8 of][]{Guidorzi2015}. However, this model is
phenomenological and only applicable for $1< \alpha_{\Delta t}\leq
3$. We find that the occurrence rate model $\lambda(t) = w
t^{-\gamma}$ has a broader application. Making a
substitution $x=wt^{-\gamma}\Delta t$, equation (\ref{pdfmod})
becomes
\begin{eqnarray}
P(\Delta t)={-w^{1/\gamma}\gamma^{-1}\Delta t^{1/\gamma-2}}\int^{x_{\rm max}}_{x_{\rm min}}x^{1-1/\gamma} e^{-x}d(x)\propto \sim\Delta t^{1/\gamma-2},
\label{modelext}
\end{eqnarray}
which implies $\alpha_{\Delta t}=2-1/\gamma$. The index $\gamma$
might vary in different astrophysical phenomena. Therefore, different
power-law indices $p$ of WTD are expected.
The case of a similar form of $\gamma\leq0$ has been partially
studied \citep{Aschwanden2010,Aschwanden2011}, where the case of
$\gamma=0$ corresponds to the stationary Poisson process and for
$\gamma\leq-1$, the corresponding power-law index of WTD is $2\leq
\alpha_{\Delta t}\leq 3$.

The power-law indices of the WTDs in astrophysics
\citep{Aschwanden2011} or geophysics, such as earthquakes
\citep{Bak2002}, are generally in the range $1\leq \alpha_{\Delta
t}\leq3$. Therefore, these WTDs in different systems, can be
understood by the form of $\lambda\propto t^{-\gamma}$ by varying
$\gamma\geq1$ or $\gamma\leq-1$. For examples, the WTD of X-ray
solar flares observed by the Hard X-Ray Spectrometer shows a
power-law with index $\alpha_{\Delta t}=0.75\pm0.1$
\citep{Pearce1993}, which can be hardly explained by previous
models. However, our model can explain it by adopting $\gamma\approx
0.8$.

\section{Summary and discussions}

In summary, we have found that the fluence, peak counts, rise time
and duration can be described by power-law distributions. This
suggests that type I X-ray bursts are governed by a SOC process,
which describes a critical state in a non-linear energy dissipation
system. For type I X-ray bursts, the critical state is reached at
$t\geq t_0$ when the nuclear energy generation rate exceeds the
diffusion loss rate. Once a burst is triggered, the nuclear energy
is released unstably. The WTD is well explained by a non-stationary
Poisson process with occurrence rate $\lambda(t)\propto \Delta
M^{-1}\propto t^{-1}$. Using very similar models, we can also
explain the WTDs in other SOC phenomena, such as various solar
phenomena and activities in flare stars. Using occurrence rate
models $\lambda(t)\propto t^{-\gamma}$ with different $\gamma$,
 we find that the power-law index of waiting time
distribution is $\alpha_{\Delta t}=2-1/\gamma$.

Meanwhile, it's noteworthy that the burst size would also follow a
power-law distribution as predicted by the SOC theory
\citep{Bak1987,Aschwanden2014}. This implies that only a part of the
fuel contributes to the observed burst. Under the circumstance, the
burst depth, area, and volume could be different to different
bursts. Interestingly, the hot spot model can provide a simple
explanation of the burst oscillations, which have been found in many
bursts \citep{Strohmayer1996,Watts2012}. In addition, the unburned
materials could participate in the subsequent burst, if the amount
of unburned material is large enough, the waiting time of the
subsequent burst could be very short, which is in agreement with a
recently study by \cite{Keek2017} using one-dimension simulations.

\section*{Acknowledgements} We have greatly benefited from the on-line
catalogue of Galloway et al. (2008). We thank the anonymous referee
and Kinwah Wu for useful suggestions. This work is supported by the
National Basic Research Program (``973" Program) of China (grant No.
2014CB845800) and National Natural Science Foundation of China
(grants 11422325, 11373022, and 11573014), and the Excellent Youth
Foundation of Jiangsu Province (BK20140016). JSW is also partially
supported by CSC.


\newpage
\begin{table}
\small
\begin{center}
    \vspace{2mm}
  \begin{tabular}{@{}c|c|c|c|c@{}}
  \hline
  $\textsf{Name}$ &  $\textsf{4U 1636-536}$ & $\textsf{EXO 0748-676}$ & $\textsf{2E 1742.9-2929}$ & $\textsf{4U 1728-34}$\\
 \hline
  $\dot m(\dot{m}_{\rm Edd})^a$ & $0.02-0.15^b$ & $7.5\pm1.8\times10^{-3}$ & $-$ & $0.041\pm0.013$\\
    $\sf{Burst~case}^c$ & case $1/2$ & case 3 & $-$ & case 2\\
    $\sf{Composition}^c$ & H\&He/He &  H\&He &  H\&He& He\\
 \hline
    $\sf{Number~of~bursts}$ & $171$ & $93$ & $83$ & $105$\\
    $\gamma$ & $1.02\pm0.11$ & $0.88\pm0.08$ & $1.21\pm0.17$ & $1.20\pm0.09$ \\
    $t_0(10^2$\,s)& $4\pm4$ & $5.3\pm5.3$ & $0.25\pm0.25$ & $<25$\\
    $T(10^6$\,s) &  $2.0\pm0.9$ & $>3.6$  & $>3$ & $>4.4$ \\
 \hline
\end{tabular}
\end{center}
\label{tab}
\protect\caption{The best-fit parameters of WTD, the accretion rate and the
burst case for four most frequent sources. $^{a}$The accretion rate
is adopted from \protect\cite{Galloway2008}, and $\dot{m}_{\rm Edd}$ is the Eddington
accretion rate. $^{b}$A range of the accretion rate indicates its
evolution with time during bursts. $^{c}$The burst type is classified by accretion rates
\protect\citep{Fujimoto1981,Narayan2003,Woosley2004,Galloway2008}: case 3
bursts are unstable mixed H/He burning with very low accretion rate
($<\sim0.01\dot{m}_{\rm Edd}$), case 2 bursts are the pure He burst
with accretion rate generally in the range around 0.01-0.1 $\dot{m}_{\rm Edd}$,
and case 1 bursts also are composed of mixed H/He with typical accretion rate 0.1-1
$\dot{m}_{\rm Edd}$.}
\end{table}


\bsp
\label{lastpage}

\end{document}